\documentclass[review,10pt,fleqn]{elsarticle}
\usepackage{amsmath,amsthm,amsfonts,amssymb,mathtools,lineno,float}
\usepackage{graphicx}
\usepackage{placeins}
\usepackage[hidelinks]{hyperref}
\usepackage[margin=0.8in]{geometry}
\hypersetup{colorlinks, linkcolor=blue, citecolor=blue}
\biboptions{sort&compress}
\newcommand{\kam}{\!\!\!}
\newcommand{\rmd}{\mathrm{d}}

\newtheorem*{pontryagin*}{Pontryagin Maximum Principle}
\journal{arXiv}
\begin{document}

\begin{frontmatter}
\title{Optimal control in cancer immunotherapy by the application of particle swarm optimization}
%%%%%%%%%%%%%%%%%%%%%%%%%%%%%%%%%%%%%%%%%%%%%%%%%%%%%%%%%%%%%%%%
%%%%%%%%%%%%%%%%%%%%%%%%%%%%%%%%%%%%%%%%%%%%%%%%%%%%%%%%%%%%%%%%
\author{Sima Sarv Ahrabi \corref{cor1}}
\cortext[cor1]{sima.sarvahrabi@sbai.uniroma1.it}
\address{Dipartimento di Scienze di Base e Applicate per l'Ingegneria, Sapienza Universit\`{a} di Roma, Via Antonio Scarpa n. 16, 00161 Rome, Italy}
%%%%%%%%%%%%%%%%%%%%%%%%%%%%%%%%%%%%%%%%%%%%%%%%%%%%%%%%%%%%%%%%
%%%%%%%%%%%%%%%%%%%%%%%%%%%%%%%%%%%%%%%%%%%%%%%%%%%%%%%%%%%%%%%%
\begin{abstract}
In this article, a well-known mathematical model of cancer immunotherapy is discussed and used to represent therapeutic protocols for cancer treatment. The optimal control problem is formulated based on the Pontryagin maximum principle to deal with adoptive cellular immunotherapy, then the problem has been solved by the application of particle swarm optimization (PSO) in combination with regular methods of solutions to optimal control problems. The results are compared with those of other researchers. It is explained how the PSO algorithm could be enlisted to obtain the optimal controls, then the obtained optimal controls are demonstrated to be more appropriate to the elimination of cancer cells by using fewer amounts of external sources of medicines.
\end{abstract}
%%%%%%%%%%%%%%%%%%%%%%%%%%%%%%%%%%%%%%%%%%%%%%%%%%%%%%%%%%%%%%%%
%%%%%%%%%%%%%%%%%%%%%%%%%%%%%%%%%%%%%%%%%%%%%%%%%%%%%%%%%%%%%%%%
\begin{keyword}
Cancer immunotherapy \sep ACI therapy \sep Optimal control \sep Pontryagin maximum principle \sep Particle swarm optimization.
%%%%%%%%%%%%%%%%%%%%%%%%%%%%%%%%%%%%%%%%%%%%%%%%%%%%%%%%%%%%%%%%
%%%%%%%%%%%%%%%%%%%%%%%%%%%%%%%%%%%%%%%%%%%%%%%%%%%%%%%%%%%%%%%%
\MSC 49J15 \sep 49J30.
%%%%%%%%%%%%%%%%%%%%%%%%%%%%%%%%%%%%%%%%%%%%%%%%%%%%%%%%%%%%%%%%
%%%%%%%%%%%%%%%%%%%%%%%%%%%%%%%%%%%%%%%%%%%%%%%%%%%%%%%%%%%%%%%%
\end{keyword}
\end{frontmatter}
%\linenumbers
%%%%%%%%%%%%%%%%%%%%%%%%%%%%%%%%%%%%%%%%%%%%%%%%%%%%%%%%%%%%%%%%%%%%%%%%%%%%
%%%%%%%%%%%%%%%%%%%%%%%%%%%%%%%%%%%%%%%%%%%%%%%%%%%%%%%%%%%%%%%%%%%%%%%%%%%%
%%%%%%%%%%%%%%%%%%%%%%%%%%%%%%%%%%  INTRODUCTION  %%%%%%%%%%%%%%%%%%%%%%%%%%
%%%%%%%%%%%%%%%%%%%%%%%%%%%%%%%%%%%%%%%%%%%%%%%%%%%%%%%%%%%%%%%%%%%%%%%%%%%%
%%%%%%%%%%%%%%%%%%%%%%%%%%%%%%%%%%%%%%%%%%%%%%%%%%%%%%%%%%%%%%%%%%%%%%%%%%%%
\section{Introduction}\label{sec.intro}

In recent years, pioneering research has been undertaken into the cancer immunotherapy as a method of enhancing the
features of cancer treatment \cite{rosenberg86,kirschner98,rosenberg04,rosenberg14,tran17}. Immunotherapy has been considered as one of the most effective methods of dealing reasonably with cancer by reinforcing humans' natural defenses in order to cope with cancer. Immunotherapy refers to the use of natural and synthetic substances to boost the immune response. The way that immunotherapy functions is as follows:
\begin{itemize}
  \item restraining or decreasing the growth of tumours.
  \item preventing cancer cells from spreading to adjoining organs.
  \item increasing the immune system's capability to destroy cancer cells.
\end{itemize}

Over the past years, the dynamics of tumour-immune interaction have been investigated, resulting in the development of various theoretical models in order to analyse the influence of immune system and tumour on each other \cite{kirschner98,kuznet94,adam96,deBoer85,baner08,castiglione06}. In \cite{eftim11} the authors have widely discussed more mathematical models, incorporating systems of delay and stochastic differential equations. The authors of \cite{kirschner98} have represented a mathematical model of immunotherapy, which is consisted of a system of three ordinary differential equations. The system, which is mentioned as the Kirschner-Panetta model, expresses the interaction between tumour cells, effector cells and IL-2, where the model will be briefly described in Section \ref{sec.KP}. The constant parameter $s_1$ and $s_2$ (see Eq. \eqref{eq.KP}), which denote respectively the external sources of effector cells and IL-2, have been considered in the model to stimulate the immune system. Since the main goal in immunotherapy is to remove the tumour cells with the least probable medication side effects, an advanced version of the model may include a time dependent external sources of medical treatment, meaning that the parameter $s_1$ and $s_2$ could be considered as control functions of time and therefore the optimum use of medical sources can be evaluated in order to achieve the optimal measure of an objective function (the so-called \textit{performance index}). Thus the main goal, the elimination of cancer cells by using the minimum amount of medical sources, can be expressed in terms of an optimal control problem.

Burden et al. \cite{burden04} have investigated the Kirschner-Panetta model by using optimal control theory in order to examine under what circumstances the tumor could be removed. They have considered a single ACI therapy in which there is not an external source of IL-2. In \cite{ghaffari10}, the authors have presented an optimal ACI therapy for the same model by making a slight modification to the performance index considered in \cite{burden04}. Then, the results were compared with those of the article \cite{burden04}. In Section \ref{sec.ACI} the optimal control problem, which has been represented in \cite{ghaffari10}, will be dealt with by utilizing the PSO algorithm, then it is demonstrated that the obtained results are more appropriate than those represented in \cite{ghaffari10}. The PSO algorithm is one of the most noticeable features of the field of nature-inspired metaheuristics. The PSO deals with an optimization problem by iteratively trying to improve the solutions. PSO was originally introduced in \cite{kennedy95,shi98modified}. In the PSO algorithm, an imaginary population of particles is defined. The particles move in the search space. Each particle represents a solution for the optimization problem. any particle is able to compare its position and solution with the best solutions and the best positions of the neighbour particles. All particles are controlled to move towards the best position and then update their current solutions. This process is reiterated until the best solution will be obtained. The PSO algorithm has experienced many changes since its introduction in \cite{kennedy95}. Many research has been conducted on the theoretical effects of the various parameters and aspects of the algorithm. More detailed studies of various aspects of the PSO algorithm could be observed in \cite{kennedy06neighborhood,kennedy02population,mendes04population,mendes04fully,ying06enhanced,bonyadi17particle,alam14research}

An optimal multi-therapeutic schedule for the administration of effectors and IL-2 has been determined in \cite{cappuccio07}. The control function $u\left(t\right)$ has been defined as the sum of $h$ separate injections, where each injection is based on the Dirac delta function. An optimization problem has been constructed and then, solved numerically by using gradient descent method. The result shows a better condition in comparison with the untreated case, nonetheless, the optimal schedule leads to an oscillation of cancer cells (see \cite[Fig. 2 (c)]{cappuccio07}). More optimal control of cancer immunotherapy could be observed in, for instance, \cite{castiglione06,castiglione07,fister05,khajanchi15}.

%%%%%%%%%%%%%%%%%%%%%%%%%%%%%%%%%%%%%%%%%%%%%%%%%%%%%%%%%%%%%%%%%%%%%%%%%%%%
%%%%%%%%%%%%%%%%%%%%%%%%%%%%%%%%%%%%%%%%%%%%%%%%%%%%%%%%%%%%%%%%%%%%%%%%%%%%
%%%%%%%%%%%%%%%%%%%%%%%%%%%%%%%%%%  SECTION 2  %%%%%%%%%%%%%%%%%%%%%%%%%%%%%
%%%%%%%%%%%%%%%%%%%%%%%%%%%%%%%%%%%%%%%%%%%%%%%%%%%%%%%%%%%%%%%%%%%%%%%%%%%%
%%%%%%%%%%%%%%%%%%%%%%%%%%%%%%%%%%%%%%%%%%%%%%%%%%%%%%%%%%%%%%%%%%%%%%%%%%%%
\section{Kirschner-Panetta model}\label{sec.KP}

In \cite{kirschner98}, the authors have introduced a mathematical model consisted of a system of three ordinary differential equations, which presents richly the immune-tumour dynamics. The Kirschner-Panetta model (KP model) indicates the immune-tumour dynamics by defining three populations, namely the effector cells such as cytotoxic T-cells, $x\left(t\right)$, the tumour cells, $y\left(t\right)$ and the concentration of IL-2, $z\left(t\right)$:
%%%%%%%%%%%%%%%%%%%%%%%%%%%%%%%%%%%%%%%%%%%%%%%%%%%%%%%%%%%%%%%%%%%%%%%%%%%%
%%%%%%%%%%%%%%%%%%%%%%%%%%%%%%%%%%%%%%%%%%%%%%%%%%%%%%%%%%%%%%%%    EQUATION
\begin{subequations}\label{eq.KP}
  \begin{align}
    \frac{\mathrm{d}x}{\mathrm{d}t}&=cy-{{\mu }_{2}}x+\frac{{p}_{1}xz}{{g}_{1}+z}+s_1\,, \label{eq.KPone}\\
    \frac{\mathrm{d}y}{\mathrm{d}t}&=r_2y\left( 1-by \right)-\frac{axy}{{g}_{2}+y}\,, \label{eq.KPtwo} \\
    \frac{\mathrm{d}{{z}}}{\mathrm{d}t}&=\frac{{p}_{2}xy}{{g}_{3}+y}-{\mu }_{3}z+s_2\,, \label{eq.KPthree}
  \end{align}
\end{subequations}
with the initial conditions
%%%%%%%%%%%%%%%%%%%%%%%%%%%%%%%%%%%%%%%%%%%%%%%%%%%%%%%%%%%%%%%%%%%%%%%%%%%%
%%%%%%%%%%%%%%%%%%%%%%%%%%%%%%%%%%%%%%%%%%%%%%%%%%%%%%%%%%%%%%%%    EQUATION
\begin{equation}\label{eq.KPinitial}
  x\left(0\right)=x_0 \,, \quad y\left(0\right)=y_0 \,, \quad z\left(0\right)=z_0 \,.
\end{equation}
The stimulation of effectors, $x$, is represented by the first and third term on the right-hand side of Eq. \eqref{eq.KPone}, where the parameter $c$ denotes the immunogenicity of the tumour, i.e.\ the ability of the tumour to provoke an immune response and the third term (Michaelis-Menten kinetics) shows that effector cells are stimulated by IL-2. The parameter $s_1$ represents an external source of effectors such as LAK cells as medical treatments. $\mu_2$ indicates the decay rate of the effectors. Eq. \eqref{eq.KPtwo} expresses the rate of change of tumour cells which is described by logistic growth $r_2y\left( 1-by \right)$ and the second term, the Michaelis-Menten term, shows the lose of tumour cells. Eq. \eqref{eq.KPthree} describes the rate of change of IL-2, produced as the result of interaction between effectors and tumour cells. The parameter $\mu_3$ denotes the decay rate of IL-2. The values of all parameters \cite{kirschner98} in Eq. \eqref{eq.KP} are given in Table \ref{tableValues}, where the units  are in $day^{-1}$ except for $b$, $g_1$, $g_2$ and $g_3$ whose units are volume.
%%%%%%%%%%%%%%%%%%%%%%%%%%%%%%%%%%%%%%%%%%%%%%%%%%%%%%%%%%%%%%%%%%%%%%%%%%%%
%%%%%%%%%%%%%%%%%%%%%%%%%%%%%%%%%%%%%%%%%%%%%%%%%%%%%%%%%%%%%%%%%%%    TABLE
\begin{table}
\centering
\caption{Values of parameters in Eq. \eqref{eq.KP}}
\begin{tabular}{ |p{3.0cm}|p{3.0cm}|p{3.0cm}|  }
 \hline
 \multicolumn{3}{|c|}{\textbf{Values of parameters}} \\
 \hline
 Equation \eqref{eq.KPone}         &    Equation \eqref{eq.KPtwo}      &     Equation \eqref{eq.KPthree}\\
  \hline
 $0\le c\le 0.05$\hphantom{=}             &    ${r}_{2}=0.18$\hphantom{=}            &      ${p}_{2}=5$\hphantom{=}        \\
 ${\mu }_{2}=0.03$\hphantom{=}            &    $b=1\times {10}^{-5}$\hphantom{=}     &      ${g}_{3}=0.1$\hphantom{=}      \\
 ${p}_{1}=0.1245$\hphantom{=}             &    $a=1$\hphantom{=}                     &      ${\mu }_{3}=10$\hphantom{=}    \\
 ${g}_{1}=2\times {10}^{3}$\hphantom{=}   &    ${g}_{2}=10$\hphantom{=}              &                                     \\
 \hline
\end{tabular}
\label{tableValues}
\end{table}

A detailed stability analysis of the KP model has been discussed in the original article \cite{kirschner98}. In immunotherapy, where external source of effector cells is used, for tumour with any immunogenicity, $c$, the effect of ACI therapy could results in elimination of tumour cells, while the level of administration of external effector cells is almost high (more than a critical level $s_{1,cr}=\frac{r_2 g_2 \mu_2}{a}$). Nevertheless, for tumours with smaller immunogenicity, the ACI treatment must be started when the tumour is still in the early stages.

%%%%%%%%%%%%%%%%%%%%%%%%%%%%%%%%%%%%%%%%%%%%%%%%%%%%%%%%%%%%%%%%%%%%%%%%%%%%
%%%%%%%%%%%%%%%%%%%%%%%%%%%%%%%%%%%%%%%%%%%%%%%%%%%%%%%%%%%%%%%%%%%%  FIGURE
%\begin{figure}[ht!]
%\centering
%\includegraphics[scale=.5]{comp2}
%\caption{Comparison of the West Function (WF) expressed in Eq. \eqref{eq.Solution} and the numerical integration of the FLE (Eq. \eqref{eq.FracLogisticKOne}), for  $\beta = 0.9$ and $u_0=0.75$\,.}
%\label{fig.KPuntreated}
%\end{figure}

%%%%%%%%%%%%%%%%%%%%%%%%%%%%%%%%%%%%%%%%%%%%%%%%%%%%%%%%%%%%%%%%%%%%%%%%%%%%
%%%%%%%%%%%%%%%%%%%%%%%%%%%%%%%%%%%%%%%%%%%%%%%%%%%%%%%%%%%%%%%%%%%%%%%%%%%%
%%%%%%%%%%%%%%%%%%%%%%%%%%%%%%%%%%  SECTION 3  %%%%%%%%%%%%%%%%%%%%%%%%%%%%%
%%%%%%%%%%%%%%%%%%%%%%%%%%%%%%%%%%%%%%%%%%%%%%%%%%%%%%%%%%%%%%%%%%%%%%%%%%%%
%%%%%%%%%%%%%%%%%%%%%%%%%%%%%%%%%%%%%%%%%%%%%%%%%%%%%%%%%%%%%%%%%%%%%%%%%%%%
\section{Adoptive cellular immunotherapy}\label{sec.ACI}

In \cite{ghaffari10}, the authors have represented an optimal control problem for the KP model
%%%%%%%%%%%%%%%%%%%%%%%%%%%%%%%%%%%%%%%%%%%%%%%%%%%%%%%%%%%%%%%%%%%%%%%%%%%%
%%%%%%%%%%%%%%%%%%%%%%%%%%%%%%%%%%%%%%%%%%%%%%%%%%%%%%%%%%%%%%%%    EQUATION
\begin{subequations}\label{eq.ACI.KP}
  \begin{align}
    \frac{\mathrm{d}x}{\mathrm{d}t}&=cy-{{\mu }_{2}}x+\frac{{p}_{1}xz}{{g}_{1}+z}+s_1 u\left(t\right)\,, \label{eq.KP.ACIone}\\
    \frac{\mathrm{d}y}{\mathrm{d}t}&=r_2y\left( 1-by \right)-\frac{axy}{{g}_{2}+y}\,, \label{eq.KP.ACItwo} \\
    \frac{\mathrm{d}{{z}}}{\mathrm{d}t}&=\frac{{p}_{2}xy}{{g}_{3}+y}-{\mu }_{3}z\,, \label{eq.KP.ACIthree}
  \end{align}
\end{subequations}
with the initial conditions
%%%%%%%%%%%%%%%%%%%%%%%%%%%%%%%%%%%%%%%%%%%%%%%%%%%%%%%%%%%%%%%%%%%%%%%%%%%%
%%%%%%%%%%%%%%%%%%%%%%%%%%%%%%%%%%%%%%%%%%%%%%%%%%%%%%%%%%%%%%%%    EQUATION
\begin{equation}\label{eq.ACI.InitialState}
  x\left(0\right)=1 \,, \quad y\left(0\right)=1 \,, \quad z\left(0\right)=1 \,.
\end{equation}
The existence of solution to Eq. \eqref{eq.ACI.KP} can be observed by referring to \cite[Theorem 9.2.1]{lukes82}. The control function $u\left( t\right)$ denotes the percentage of the external source of effectors, $s_1$, and therefore it is bounded by $0 \le u\left( t\right) \le 1$. The goal is to design a protocol leading to eradication of cancer cells, $y\left( t\right)$, at the end of treatment, in addition to using the least amount of drug, and causing the level of cancer cells to be as low as possible and the levels of effectors and IL-2 to be remained at a high degree. Thus the performance index, which must be maximized, is obtained as follows:
%%%%%%%%%%%%%%%%%%%%%%%%%%%%%%%%%%%%%%%%%%%%%%%%%%%%%%%%%%%%%%%%%%%%%%%%%%%%
%%%%%%%%%%%%%%%%%%%%%%%%%%%%%%%%%%%%%%%%%%%%%%%%%%%%%%%%%%%%%%%%    EQUATION
\begin{equation}\label{eq.ACI.Performance}
  J\left(u\right) = -Ay\left( {t}_{f} \right)+\int_{0}^{t_f} \kam {\left( x\left( t \right)-y\left( t \right)+z\left( t \right)-\frac{1}{2}B{{\left( u\left( t \right) \right)}^{2}} \right)\rmd t} \,,
\end{equation}
where ${t}_{f}$ is the fixed final time and denotes the period of the treatment. The linear payoff term $-Ay\left(t_f\right)$ is considered in order to minimize the level of tumour cells at the end of treatment, where $A$ is a constant coefficient. The coefficient $B$ in quadratic term $-\frac{1}{2}B{{\left( u\left( t \right) \right)}^{2}}$ expresses the importance of minimizing $u$ in comparison to minimization of tumour cells. The terme $x\left( t \right)-y\left( t \right)+z\left( t \right)$ is added to Eq. \eqref{eq.ACI.Performance} with the intention of holding the cancer cells at a low level, and the effectors and IL-2 as high as possible during the treatment. The existence and uniqueness of an optimal control has been proved in \cite{burden04}. The Pontryagin maximum principle \cite{pontryagin87} is used to formulate the necessary conditions of optimality. The Hamiltonian function \cite{kirk12} is as follows
%%%%%%%%%%%%%%%%%%%%%%%%%%%%%%%%%%%%%%%%%%%%%%%%%%%%%%%%%%%%%%%%%%%%%%%%%%%%
%%%%%%%%%%%%%%%%%%%%%%%%%%%%%%%%%%%%%%%%%%%%%%%%%%%%%%%%%%%%%%%%    EQUATION
\begin{eqnarray}\label{eq.ACI.Hamiltonian}
  H & = & x-y+z-\frac{1}{2}B u^2 + \lambda_1 \left(  cy-{{\mu }_{2}}x+\displaystyle \frac{{{p}_{1}}xz}{{{g}_{1}}+z}+{{s}_{1}} u\left(t\right)  \right) \nonumber    \\
  &   & + \lambda_2 \left(  {{r}_{2}}y\left( 1-by \right)-\displaystyle \frac{axy}{{{g}_{2}}+y} \right)+\lambda_3 \left(\displaystyle \frac{{{p}_{2}}xy}{{{g}_{3}}+y}-{{\mu }_{3}}z\right)\,,
\end{eqnarray}
where $\lambda_1$, $\lambda_2$ and $\lambda_3$ are the adjoint variables, and the adjoint system is:
%%%%%%%%%%%%%%%%%%%%%%%%%%%%%%%%%%%%%%%%%%%%%%%%%%%%%%%%%%%%%%%%%%%%%%%%%%%%
%%%%%%%%%%%%%%%%%%%%%%%%%%%%%%%%%%%%%%%%%%%%%%%%%%%%%%%%%%%%%%%%    EQUATION
\begin{subequations}\label{eq.ACI.Adjoint}
  \begin{align}
    \displaystyle \frac{\mathrm{d}\lambda_1}{\mathrm{d}t} = & - \left[ 1 + \lambda_1 \left( -{\mu }_{2}+\displaystyle \frac{{{p}_{1}}z}{{{g}_{1}}+z} \right) - \displaystyle \frac{ay}{{{g}_{2}}+y}\lambda_2 + \displaystyle \frac{{{p}_{2}}y}{{{g}_{3}}+y}\lambda_3 \right]\,, \label{eq.ACI.Adjointone}\\
    \displaystyle \frac{\mathrm{d}\lambda_2}{\mathrm{d}t} = & - \left[ -1 + c\lambda_1 + \left( r_2-2r_2by \right)\lambda_2 - \displaystyle \frac{a{g}_2x}{{\left(g_2+y\right)}^2} \lambda_2 + \displaystyle \frac{{p}_2{g}_3x}{{\left(g_3+y\right)}^2} \lambda_3  \right]\,, \label{eq.ACI.Adjointtwo} \\
    \displaystyle \frac{\mathrm{d}\lambda_3}{\mathrm{d}t} = & - \left[ 1 + \displaystyle \frac{{p}_1{g}_1x}{{\left(g_1+z\right)}^2} \lambda_1 -\mu_3 \lambda_3 \right]\,. \label{eq.ACI.Adjointthree}
  \end{align}
\end{subequations}
The values of adjoint variables at the final time $t_f$ are evaluated by using the transversality condition \cite[Eq. 5.1-20]{lukes82}:
%%%%%%%%%%%%%%%%%%%%%%%%%%%%%%%%%%%%%%%%%%%%%%%%%%%%%%%%%%%%%%%%%%%%%%%%%%%%
%%%%%%%%%%%%%%%%%%%%%%%%%%%%%%%%%%%%%%%%%%%%%%%%%%%%%%%%%%%%%%%%    EQUATION
\begin{equation}\label{eq.ACI.FinalAdjoint}
  \lambda_1 \left( t_f \right) =0\,, \qquad \lambda_2 \left( t_f \right) =-A\,, \qquad \lambda_3 \left( t_f \right) =0\,.
\end{equation}
The derivative of Hamiltonian \eqref{eq.ACI.Hamiltonian} with respect to $u$ is calculated as follows:
%%%%%%%%%%%%%%%%%%%%%%%%%%%%%%%%%%%%%%%%%%%%%%%%%%%%%%%%%%%%%%%%%%%%%%%%%%%%
%%%%%%%%%%%%%%%%%%%%%%%%%%%%%%%%%%%%%%%%%%%%%%%%%%%%%%%%%%%%%%%%    EQUATION
\begin{equation}\label{eq.ACI.dHdu}
  \frac{\partial H}{\partial u} = -Bu + \lambda_1 s_1 \,.
\end{equation}
In order to characterize the optimal control, there exist three cases \cite[Eq. 8.6]{lenhart07}:
\begin{enumerate}
  \item If $u\left(t\right)=0$, then $\frac{\partial H}{\partial u} \le 0$, therefore $\lambda_1 \le 0$\,.
  \item If $0<u\left(t\right)<1$, then $\frac{\partial H}{\partial u} = 0$, therefore $u=\frac{s_1}{B}\lambda1$, where $0<\lambda_1 < \frac{B}{s_1}$ \,.
  \item If $u\left(t\right)=1$, then $\frac{\partial H}{\partial u} \ge 0$, therefore $\lambda_1 \ge \frac{B}{s_1}$ \,.
\end{enumerate}
Thus, the characterization of optimality is:
%%%%%%%%%%%%%%%%%%%%%%%%%%%%%%%%%%%%%%%%%%%%%%%%%%%%%%%%%%%%%%%%%%%%%%%%%%%%
%%%%%%%%%%%%%%%%%%%%%%%%%%%%%%%%%%%%%%%%%%%%%%%%%%%%%%%%%%%%%%%%    EQUATION
\begin{equation}\label{eq.ACI.optimality}
  u\left(t\right) = \begin{cases}
                      0 &\quad \mbox{if }\; \lambda_1 \le 0 \,,\\
                      \frac{s_1}{B}\lambda1 &\quad \mbox{if }\; 0<\lambda_1 < \frac{B}{s_1} \,,\\
                      1 &\quad \mbox{if }\; \lambda_1 \ge \frac{B}{s_1} \,.
                    \end{cases}
\end{equation}
The optimal control problem, described above, consists of state system \eqref{eq.ACI.KP}, adjoint system \eqref{eq.ACI.Adjoint}, and optimality condition \eqref{eq.ACI.optimality}. This is obviously a two-point boundary value problem, since the initial state of the state variables (see \eqref{eq.ACI.InitialState}) and the final state of the adjoint variables (see \eqref{eq.ACI.FinalAdjoint}) are known. Despite the fact that the general methods of solving these types of problems are those such as the shooting method, the problem mentioned above may be solved more straightforward due to the fact that state system \eqref{eq.ACI.KP} is independent of adjoint variables. Thus, an initial guess for adjoint variables is made, by which the control function, $u\left(t\right)$, can be evaluated referring to Eq. \eqref{eq.ACI.optimality}. Using the obtained control function, $u\left(t\right)$, state system \eqref{eq.ACI.KP} is solved forward in time, then adjoint system \eqref{eq.ACI.Adjoint} is updated backward in time by using the evaluated state variables. The process continues until the differences between the current and previous values of state variables, adjoint variables, control function and the performance index are within a specified error bound.

Although solving the problem sounds uncomplicated, here the main issue is that any initial guess for the adjoint variables does not necessarily leads to finding the optimal control maximizing the performance index. In most cases, the initial guess is such that the algorithm does not even converge to a solution and more important, the convergence to the optimal control is almost impossible. As illustrated in \cite{ghaffari10}, typical optimal controls for these types of problems are generally bang-bang, in the sense that the optimal control switches periodically between lower and upper bounds. This can be observed from Eq. \eqref{eq.ACI.optimality}. While $B=5$ and $s_1 = 500$ (see, for instance, \cite[Fig. 2]{ghaffari10}), the value of $\frac{B}{s_1}$ is $0.01$ and therefore $u$ is practically a bang-bang control.

Due to above statement, a hybrid of the PSO algorithm and the method mentioned above is used to find the optimal therapeutic protocol for the problem. First the optimal control and corresponding adjoint variables are obtained by applying the PSO algorithm, where the control function, $u\left(t\right)$, is considered to be a bang-bang control. While a specified criterion is fulfilled (for instance, the convergence of the performance index), the algorithm is switched to the above method to search for a probable better solution. The problem is solved for two different set of values of the parameters $c$, $B$ and $s_1$. The period of therapy is considered to be $350$ days, i.e.\ $t_f=350$.

%%%%%%%%%%%%%%%%%%%%%%%%%%%%%%%%%%%%%%%%%%%%%%%%%%%%%%%%%%%%%%%%%%%%%%%%%%%%
%%%%%%%%%%%%%%%%%%%%%%%%%%%%%%%%%%%%%%%%%%%%%%%%%%%%%%%%%%%%%%%%%%%%%%%%%%%%
%%%%%%%%%%%%%%%%%%%%%%%%%%%%%%%%%%  SECTION 4  %%%%%%%%%%%%%%%%%%%%%%%%%%%%%
%%%%%%%%%%%%%%%%%%%%%%%%%%%%%%%%%%%%%%%%%%%%%%%%%%%%%%%%%%%%%%%%%%%%%%%%%%%%
%%%%%%%%%%%%%%%%%%%%%%%%%%%%%%%%%%%%%%%%%%%%%%%%%%%%%%%%%%%%%%%%%%%%%%%%%%%%

\section{Discussion}\label{sec.Discussion}

\begin{figure}[!ht]
\centering
\includegraphics[ height=9cm, width=13cm]{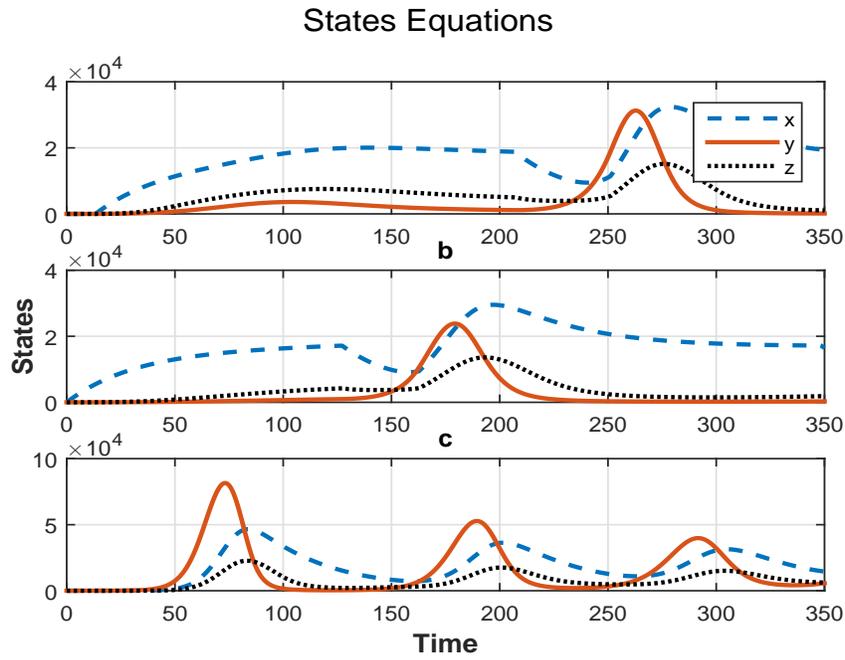}
\caption{State variables, (a): in this work, (b): in \cite{ghaffari10}, (c): in \cite{kirschner98} for $c=0.04$, $s1=500$, $B=1$.}
\label{state4500}
\end{figure}

\begin{figure}[!ht]
\centering
\includegraphics[ height=9cm, width=13cm]{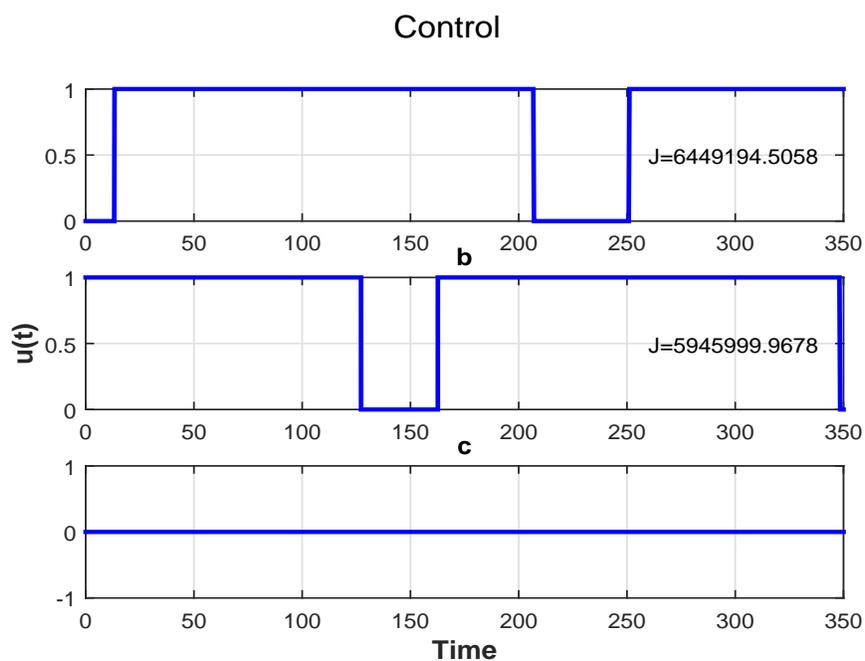}
\caption{Optimal Control, (a): in this work, (b): in \cite{ghaffari10}, (c): in \cite{kirschner98} for $c=0.04$, $s1=500$, $B=1$.}
\label{control4500}
\end{figure}

The results are obtained for three different cases, based on the choice of different values for $c$, $s_1$, and $B$. The duration of therapy is consider to be $350$ days, i.e.\ $t_f = 350$:

\textbf{Case 1:} $c=0.04$, $s_1=500$, and $B=1$. Fig. \ref{state4500} shows the state variables, i.e.\ tumor cells ($x$), the effector cells ($y$), and the concentration of interleukin-2 ($z$). The non-tumor equilibrium point in this case is unstable because the value of $s_1$ is smaller than critical value $s_{1,cr} =540$ \cite{kirschner98}. Nonetheless, the control pushes the system to the area with smaller cancerous cells. In this work, in comparison with the work done in \cite{ghaffari10}, the amount of total used drug has been decreased (Fig. \ref{control4500}), and the maximum value of interleukin-2 is larger. The most important thing, in this work, is that the objective function is maximized ($J=6449194$) which is larger than the objective function obtained in \cite{ghaffari10}.

\begin{figure}[!htbp]
\centering
\includegraphics[ height=9cm, width=13cm]{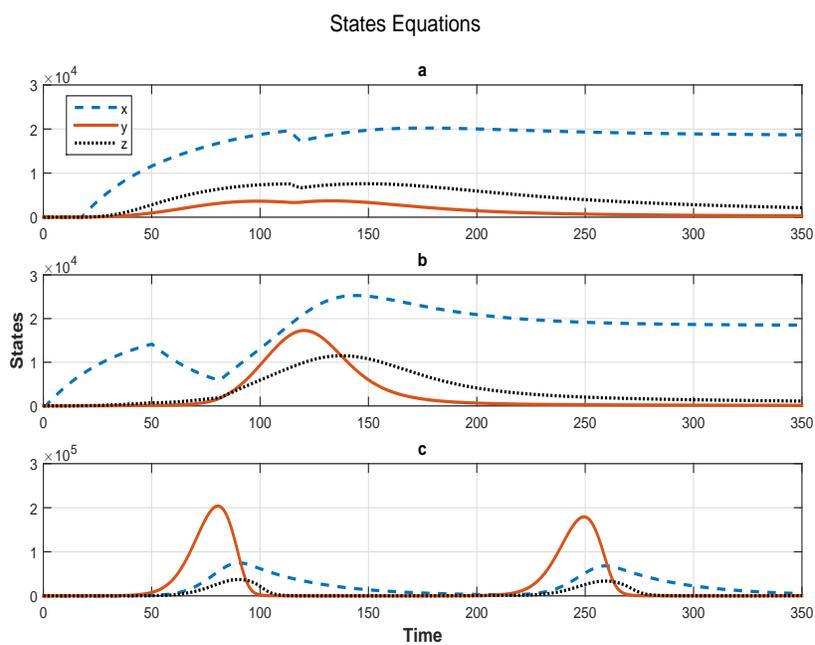}
\caption{State variables, (a): in this work, (b): in \cite{ghaffari10}, (c): in \cite{kirschner98} for $c=0.025$, $s1=550$, $B=1$.}
\label{states255501}
\end{figure}

\begin{figure}[!htbp]
\centering
\includegraphics[ height=9cm, width=13cm]{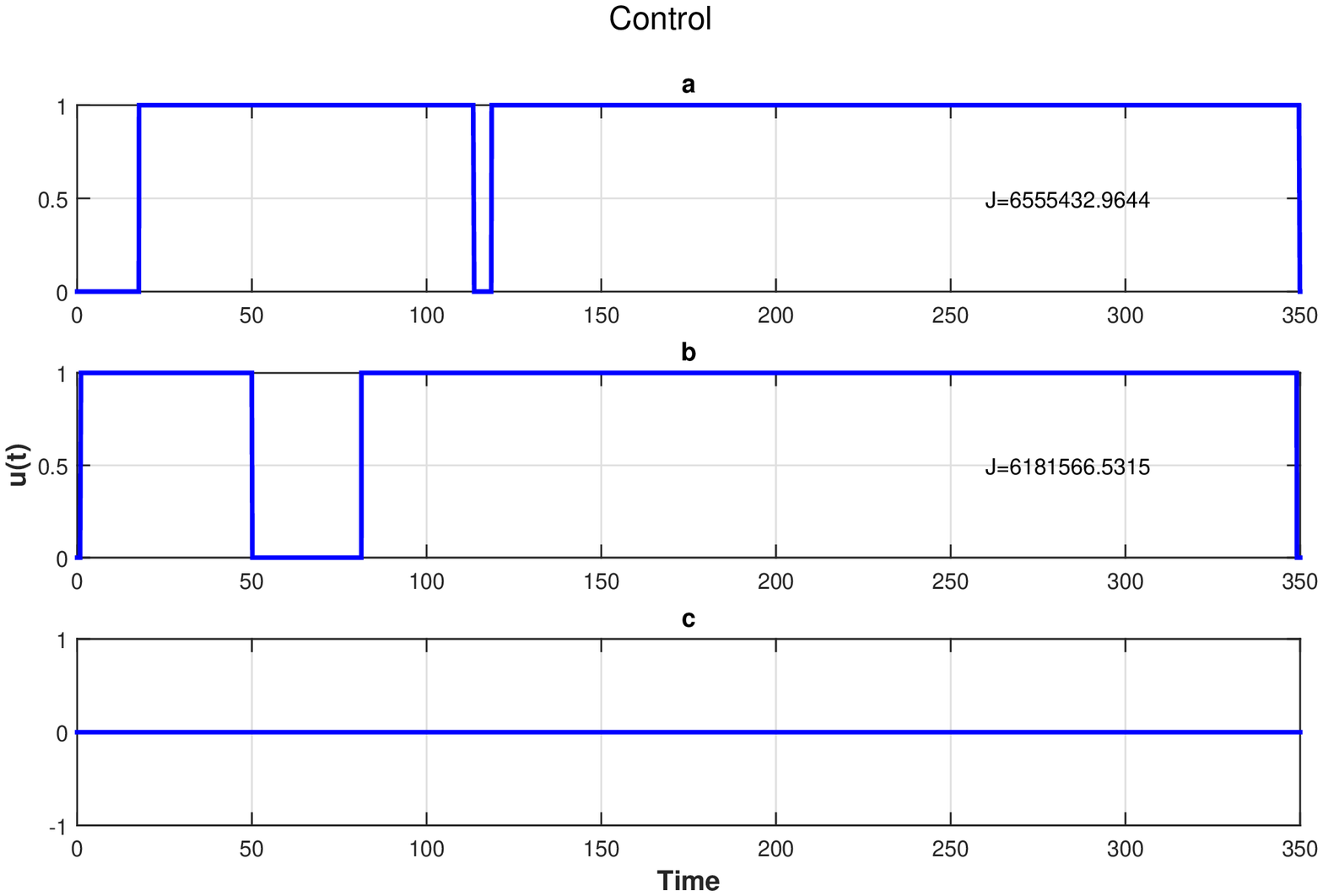}
\caption{Optimal Control, (a): in this work, (b): in \cite{ghaffari10}, (c): in \cite{kirschner98} for $c=0.025$, $s1=550$, $B=1$.}
\label{control255501}
\end{figure}

\textbf{Case 2:} $c=0.025$, $s_1=550$, and $B=1$. The results are shown in Figs. \ref{states255501} and \ref{control255501}. Since $s_1 > s_{1,cr}$, the non-tumour state is stable. Thus, it is expected that the tumor completely inhibited. Fig. \ref{states255501} shows that the maximum value of the tumour has been minimized over the treatment. In addition, as it is illustrated in Fig. \ref{control255501}, The performance index has been maximized in comparison with \cite{ghaffari10}.

\begin{figure}[!htbp]
\centering
\includegraphics[ height=9cm, width=13cm]{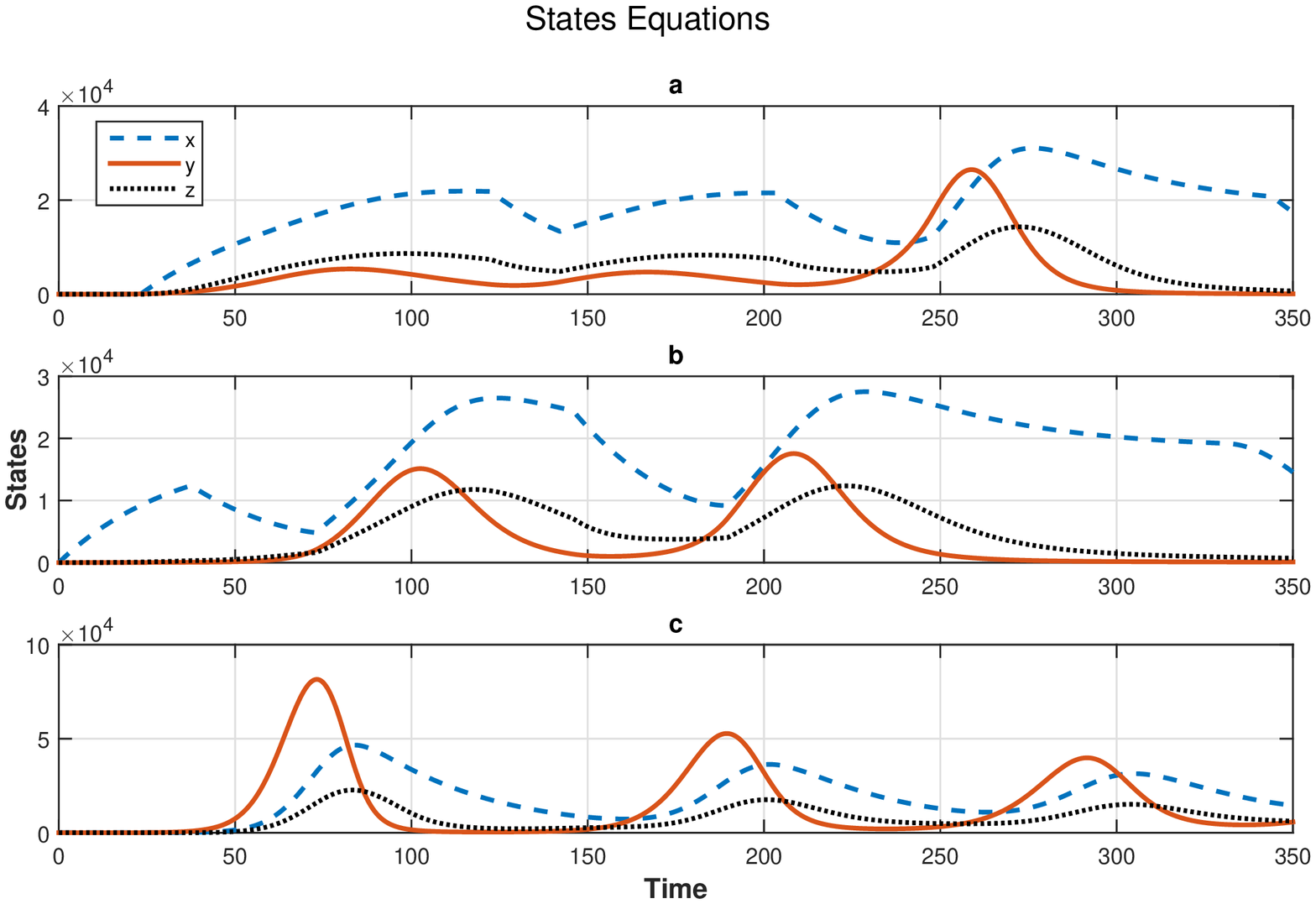}
\caption{State variables, (a): in this work, (b): in \cite{ghaffari10}, (c): in \cite{kirschner98} for $c=0.04$, $s1=550$, $B=10000$.}
\label{states455010000}
\end{figure}

\begin{figure}[!htbp]
\centering
\includegraphics[ height=9cm, width=13cm]{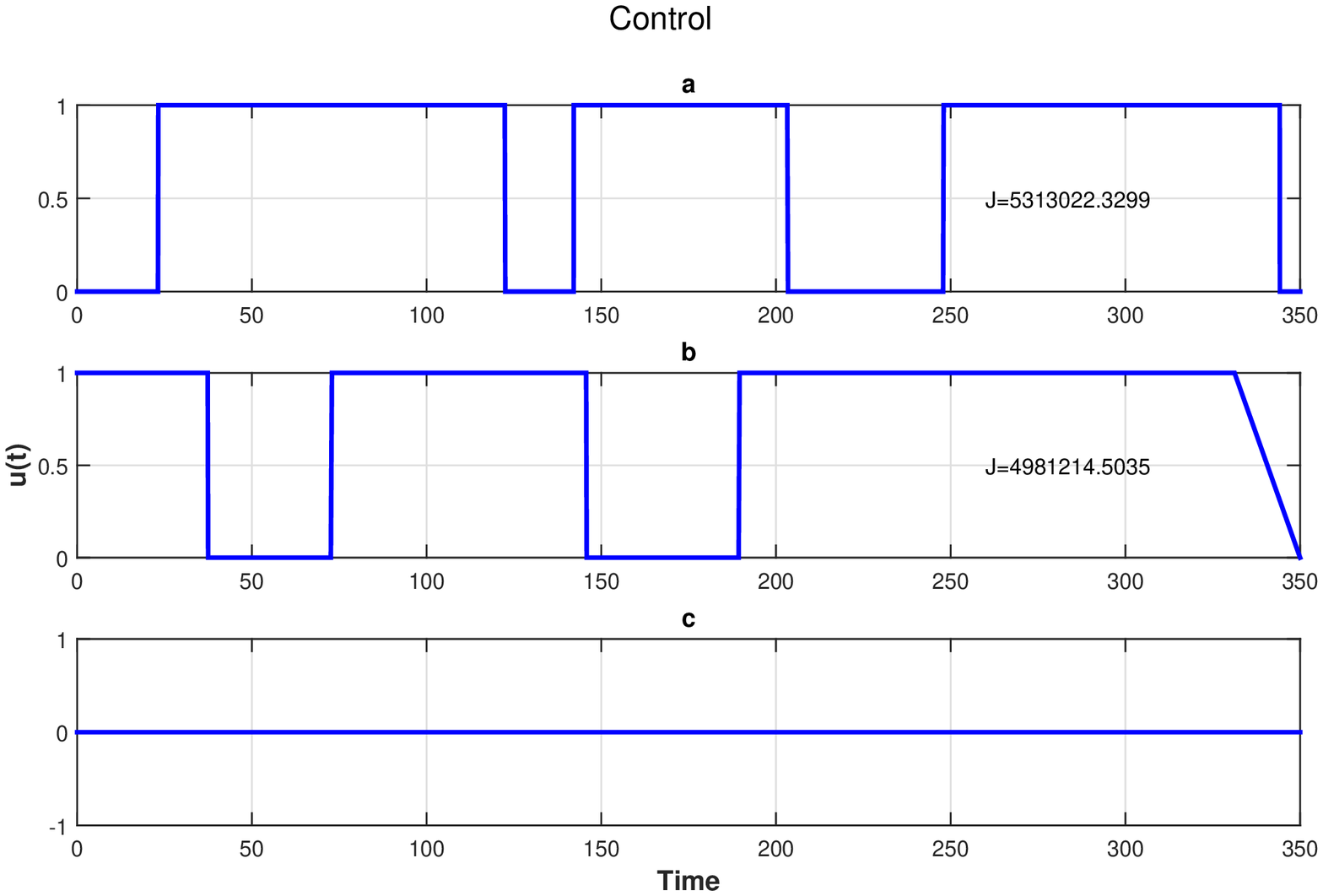}
\caption{Optimal Control, (a): in this work, (b): in \cite{ghaffari10}, (c): in \cite{kirschner98} for $c=0.04$, $s1=550$, $B=10000$.}
\label{control455010000}
\end{figure}

\textbf{Case 3:} $c=0.04$, $s_1=550$, and $B=10000$. The results are shown in Figs. \ref{states455010000} and \ref{control455010000}. Since $s_1 > s_{1,cr}$, the non-tumour state is stable. The stress is here on minimizing the total amount of administration, since the parameter $B$ has been chosen to be very large. The performance index has been maximized in comparison with the work done in \cite{ghaffari10}. The maximum value of tumour cells are at a lower level compared with \cite{ghaffari10}.
\FloatBarrier
\section{Conclusion}
As it was shown, by using the hybrid method of PSO and forward-backward sweep method the obtained results are much better and more acceptable than those represented in other research. Using the PSO algorithm along with classical methods for numerical solution of optimal controls definitely improves the performance of finding the optimal control. It is a fact that classical approaches are very time-consuming, since any initial guess cannot guarantee the convergence.
%%%%%%%%%%%%%%%%%%%%%%%%%%%%%%%%%%%%%%%%%%%%%%%%%%%%%%%%%%%%%%%%%%%%%%%%%%%%%
%%%%%%%%%%%%%%%%%%%%%%%%%%%%%%%%%%%%%%%%%%%%%%%%%%%%%%%%%%%%%%%%%%%%%%%%%%%%%
%%%%%%%%%%%%%%%%%%%%%%%%%%%%%%%%%%%%%%%%%%%%%%%%%%%%%%%%%%%%%%%%%%%%%%%%%%%%%
%%%%%%%%%%%%%%%%%%%%%%%%%%%%%%%%%%%%%%%%%%%%%%%%%%%%%%%%%%%%%%%%%%%%%%%%%%%%%
\section*{References}
\bibliographystyle{elsarticle-num}
\bibliography{references}
\end{document}